\documentclass[twocolumn,aps,prl,showpacs]{revtex4}

\begin{document}
\title{Reply to Marinatto's comment on ``Bell's
theorem without inequalities and without alignments''}
\author{Ad\'{a}n Cabello}
\email{adan@us.es}
\affiliation{Departamento de F\'{\i}sica
Aplicada II, Universidad de Sevilla, 41012 Sevilla, Spain}
\date{\today}


\begin{abstract}
Marinatto claims that in the proof of Bell's theorem without
inequalities and without alignments [A. Cabello, Phys.~Rev.~Lett.
{\bf 91}, 230403 (2003)], local observables cannot be measured by
means of tests on individual qubits. Marinatto's claim is
incorrect. To support this, the proof is explicitly rewritten in
terms of tests on individual qubits.
\end{abstract}

\maketitle


Marinatto claims~\cite{Marinatto04} that in the proof of Bell's
theorem without inequalities and without
alignments~\cite{Cabello03}, local observables cannot be measured
by means of tests on individual qubits. We believe that this claim
is incorrect. To support this, we explicitly rewrite the proof in
terms of tests on individual qubits.

Alice and Bob share eight qubits prepared in the state
$|\eta\rangle$ given by Eq.~(1) in~\cite{Cabello03}. Alice has the
first four qubits and Bob the remaining four qubits. On her four
qubits Alice can measure either $R _A \sigma_{z1}$, $R _A
\sigma_{z2}$, $R _A \sigma_{x3}$, and $R _A \sigma_{x4}$, or
${\cal R}_A \sigma_{z1}$, ${\cal R}_A \sigma_{x2}$, ${\cal R}_A
\sigma_{z3}$, and ${\cal R}_A \sigma_{x4}$, where $\sigma_{z1}$ is
the spin component of the first qubit along the $z$ direction, and
$R_A$ (and ${\cal R}_A$) is any rotation of Alice's setups for
measuring her four qubits. Analogously, on his four qubits Bob can
measure either $R _B \sigma_{z5}$, $R _B \sigma_{z6}$, $R _B
\sigma_{x7}$, and $R _B \sigma_{x8}$, or ${\cal R}_B \sigma_{z5}$,
${\cal R}_B \sigma_{x6}$, ${\cal R}_B \sigma_{z7}$, and ${\cal
R}_B \sigma_{x8}$, where $R_B$ (and ${\cal R}_B$) is any rotation
of Bob's setups for measuring his four qubits. Let us denote by
$00\bar{0}\bar{0},\ldots, 11\bar{1}\bar{1}$, the $16$ possible
outcomes of Alice (Bob) measuring $R _A \sigma_{z1}$, $R _A
\sigma_{z2}$, $R _A \sigma_{x3}$, and $R _A \sigma_{x4}$ ($R _B
\sigma_{z5}$, $R _B \sigma_{z6}$, $R _B \sigma_{x7}$, and $R _B
\sigma_{x8}$), and by $0\bar{0}0\bar{0},\ldots, 1\bar{1}1\bar{1}$,
the $16$ possible outcomes of Alice (Bob) measuring ${\cal R}_A
\sigma_{z1}$, ${\cal R}_A \sigma_{x2}$, ${\cal R}_A \sigma_{z3}$,
and ${\cal R}_A \sigma_{x4}$ (${\cal R}_B \sigma_{z5}$, ${\cal
R}_B \sigma_{x6}$, ${\cal R}_B \sigma_{z7}$, and ${\cal R}_B
\sigma_{x8}$). If Alice (Bob) obtains $01\bar{0}\bar{1}$,
$01\bar{1}\bar{0}$, $10\bar{0}\bar{1}$, or $10\bar{1}\bar{0}$, she
(he) will annotate $F_A=-1$ ($F_B=-1$) as a collective result of
her (his) four measurements. In the other $12$ possible cases, she
(he) will annotate $F_A=1$ ($F_B=1$). If Alice (Bob) obtains
$0\bar{0}1\bar{1}$, $0\bar{1}1\bar{0}$, $1\bar{0}0\bar{1}$, or
$1\bar{1}0\bar{0}$, she (he) will annotate $G_A=-1$ ($G_B=-1$) as
a collective result of her (his) four measurements. In the other
$12$ possible cases, she (he) will annotate $G_A=1$ ($G_B=1$).
With this notation, if Alice measures $R _A \sigma_{z1}$, $R _A
\sigma_{z2}$, $R _A \sigma_{x3}$, and $R _A \sigma_{x4}$, and Bob
measures $R _B \sigma_{z5}$, $R _B \sigma_{z6}$, $R _B
\sigma_{x7}$, and $R _B \sigma_{x8}$, the joint probability that,
in the state $|\eta\rangle$, Alice obtains $F_A =1$ and Bob
obtains $F_B=1$ is
\begin{equation}
P(F_A =1,F_B=1) = 0,
\label{M4}
\end{equation}
because, in the state $|\eta\rangle$, the $12^2$ joint
probabilities $P(00\bar{0}\bar{0},00\bar{0}\bar{0}),\ldots,
P(11\bar{1}\bar{1},11\bar{1}\bar{1})$ are zero.

If Alice measures $R _A \sigma_{z1}$, $R _A \sigma_{z2}$, $R _A
\sigma_{x3}$, and $R _A \sigma_{x4}$, and Bob measures ${\cal R}_B
\sigma_{z5}$, ${\cal R}_B \sigma_{x6}$, ${\cal R}_B \sigma_{z7}$,
and ${\cal R}_B \sigma_{x8}$, the probability that, in the state
$|\eta\rangle$, Alice obtains $F_A =1$, conditioned to Bob
obtaining $G_B=1$ is
\begin{equation}
P(F_A =1\,|\,G_B=1) = 1,
\label{M3}
\end{equation}
because, in the state $|\eta\rangle$, the $4 \times 12$ joint
probabilities $P(01\bar{0}\bar{1},0\bar{0}0\bar{0}),\ldots,
P(10\bar{1}\bar{0},1\bar{1}1\bar{1})$ are zero.

Analogously, if Alice measures ${\cal R}_A \sigma_{z1}$, ${\cal
R}_A \sigma_{x2}$, ${\cal R}_A \sigma_{z3}$, and ${\cal R}_A
\sigma_{x4}$, and Bob measures $R _B \sigma_{z5}$, $R _B
\sigma_{z6}$, $R _B \sigma_{x7}$, and $R _B \sigma_{x8}$, then in
the state $|\eta\rangle$,
\begin{equation}
P(F_B =1\,|\,G_A=1) = 1,
\label{M2}
\end{equation}
because, in the state $|\eta\rangle$, the $12 \times 4$ joint
probabilities $P(0\bar{0}0\bar{0},01\bar{0}\bar{1}),\ldots,
P(1\bar{1}1\bar{1},10\bar{1}\bar{0})$ are zero.

Finally, if Alice measures ${\cal R}_A \sigma_{z1}$, ${\cal R}_A
\sigma_{x2}$, ${\cal R}_A \sigma_{z3}$, and ${\cal R}_A
\sigma_{x4}$, and Bob measures ${\cal R}_B \sigma_{z5}$, ${\cal
R}_B \sigma_{x6}$, ${\cal R}_B \sigma_{z7}$, and ${\cal R}_B
\sigma_{x8}$, then in the state $|\eta\rangle$,
\begin{equation}
P(G_A=1,G_B=1) = {9 \over 112},
\label{M1}
\end{equation}
because, in the state $|\eta\rangle$, the sum of the $12^2$ joint
probabilities $P(0\bar{0}0\bar{0},0\bar{0}0\bar{0}),\ldots,
P(1\bar{1}1\bar{1},1\bar{1}1\bar{1})$ is $9/112$.

Eqs. (\ref{M4})--(\ref{M1}) allow us to develop a
Hardy-like~\cite{Hardy93} proof of Bell's theorem without
inequalities (see~\cite{Cabello03} for details), using only single
qubit measurements, and with the remarkable property that Alice
and Bob's setups do not need to be aligned, because the required
perfect correlations are achieved for any local rotation of the
setups.




\begin{thebibliography}{99}

\bibitem{Marinatto04}
L. Marinatto,
preceding Comment,
Phys. Rev. Lett. {\bf 93}, \ldots (2004).

\bibitem{Cabello03}
A. Cabello,
Phys. Rev. Lett. {\bf 91}, 230403 (2003).

\bibitem{Hardy93}
L. Hardy,
Phys. Rev. Lett. {\bf 71}, 1665 (1993).

\end{thebibliography}
\end{document}